\documentclass[sigconf]{acmart}
\usepackage{amsmath}
\usepackage{graphicx}
\usepackage{amsfonts}
\usepackage{amsthm}
\usepackage{bbm}
\usepackage{mathtools}
\usepackage{balance}
\usepackage{bm}
\usepackage{color}
\usepackage{enumerate}
\usepackage{sgame, tikz}
\usepackage{accents}
\usepackage{microtype}
\usepackage{float}
\newcommand{\Comments}{0}
\newcommand{\mynote}[2]{\ifnum\Comments=1\textcolor{#1}{#2}\fi}

\newcommand{\ignore}[1]{}

\newtheorem{prop}{Proposition}

\theoremstyle{definition}

\copyrightyear{2018}
\acmYear{2018} 
\setcopyright{iw3c2w3}
\acmConference[WWW 2018]{The 2018 Web Conference}{April 23--27, 2018}{Lyon, France}
\acmBooktitle{WWW 2018: The 2018 Web Conference, April 23--27, 2018, Lyon, France}
\acmPrice{}
\acmDOI{10.1145/3178876.3186044}
\acmISBN{978-1-4503-5639-8/18/04}
\begin{document}
\title[A Short-term Intervention for Long-term Fairness in the Labor Market]{A Short-term Intervention for Long-term\\ Fairness in the Labor Market}
\author{Lily Hu}
\affiliation{
	\institution{Harvard University}
	\city{Cambridge}
	\state{MA}}
	 \email{lilyhu@g.harvard.edu}
\author{Yiling Chen}
\affiliation{
	\institution{Harvard University}
	\city{Cambridge}
	\state{MA}}
	 \email{yiling@seas.harvard.edu}

%

\begin{abstract}
The persistence of racial inequality in the U.S. labor market against a general backdrop of formal equality of opportunity is a troubling phenomenon that has significant ramifications on the design of hiring policies. In this paper, we show that current group disparate outcomes may be immovable even when hiring decisions are bound by an input-output notion of ``individual fairness.'' Instead, we construct a dynamic reputational model of the labor market that illustrates the reinforcing nature of asymmetric outcomes resulting from groups' divergent accesses to resources and as a result, investment choices. To address these disparities, we adopt a dual labor market composed of a Temporary Labor Market (TLM), in which firms' hiring strategies are constrained to ensure statistical parity of workers granted entry into the pipeline, and a Permanent Labor Market (PLM), in which firms hire top performers as desired. Individual worker reputations produce externalities for their group; the corresponding feedback loop raises the collective reputation of the initially disadvantaged group via a TLM fairness intervention that need not be permanent. We show that such a restriction on hiring practices induces an equilibrium that, under particular market conditions, Pareto-dominates those arising from strategies that statistically discriminate or employ a ``group-blind'' criterion. The enduring nature of equilibria that are both inequitable and Pareto suboptimal suggests that fairness interventions beyond procedural checks of hiring decisions will be of critical importance in a world where machines play a greater role in the employment process.
\end{abstract}
\maketitle

\section{Introduction}
As algorithms are increasingly deployed to make social decisions that have previously been under the sole purview of humans, a growing body of work has challenged the reigning primacy of optimality and efficiency when issues of bias and discrimination are potentially at stake. Research in the growing field of algorithmic fairness has sought to address these concerns about the machine decision-making process by examining and manipulating standard tasks such as ranking or classification under generalized constraints of ``fairness.'' Such computational notions of fairness have been varied but two broad opposing perspectives have proposed solutions that either defend fairness at the individual level (similar individuals are treated similarly) \citep{dwork2012fairness} or at the group level (groups are awarded proportional representation) \citep{kamishima2011fairness, feldman2015certifying}. While this paper similarly adopts a constraint-based intervention to achieve fairness, we depart from standard accounts of fairness that consider static domain-general algorithms and instead develop a dynamic model for the specific domain of decision-making in the labor market. Our work considers the role that firms' hiring practices play in perpetuating economic inequalities between social groups by way of the disparate outcomes that groups experience in their employment opportunities and wage prospects. We address the issue by building upon a dynamic model of worker and firm behavior that has been shown to generate the asymmetric group outcomes that are observed empirically between black and white workers in the United States \cite{cain1986economic, altonji1999race, fryer2013racial} and appending a constraint on firms' hiring practices that successfully induces a group-equitable equilibrium.

As we focus on the particular domain of labor market dynamics, our paper draws upon an extensive literature in economics. The theory of statistical discrimination, originally set forth in two seminal papers by Phelps \cite{phelps1972statistical} and Arrow \cite{arrow1973theory}, explains disparate group outcomes as the result of rational agent behaviors that lock a system into an unfavorable equilibrium. In the basic model, workers compete for a skilled job with wage $w$. Skill acquisition requires workers to expend an investment cost of $c$, which is distributed according to a function $F$. A worker's investment decision is an assessment of her expected wage gain compared with her investment cost. Firms seek information about a worker's hidden \textit{ability} level but can only base hiring decisions on observable attributes: her noisy \textit{investment} signal and group membership. The firm's response to this missing information problem is to update its beliefs about a worker's qualifications by drawing on its prior for her group's ability levels. Therefore if a firm holds different priors for different groups, it will also set different group-specific hiring thresholds. Further, since these distinct thresholds are observed and internalized by workers, they adjust their own investment strategies accordingly---individuals within the unfavored group will lower their investment levels, and individuals in the favored group will continue to invest at a high level. Notably, even when the distribution of investment costs $F$ is the same for each group\footnote{This has been the standard assumption in the economics literature since Arrow \cite{arrow1973theory}.}, an asymmetric equilibrium can arise in which groups invest at different levels, further informing firms' distinct priors and reinforcing disparate employment prospects. In other words, rational workers and firms best respond in ways that exactly confirm the others' beliefs and strategies, and thus, the discriminatory outcome is ``justified.''

A proponent of ``individual fairness'' may diagnose the problem of statistical discrimination as a failure to treat candidates of similar investments similarly\footnote{In the exposition of ``individual fairness'' proposed by Dwork et al. \cite{dwork2012fairness}, the built-in flexibility of the generic similarity metric between persons can include group membership and even be used to justify ``fair affirmative action.'' However, within an economic signaling environment where firms' hiring standards affect workers' investments, a more flexible metric approach that compares quality within and across groups still fails to account for the strategy and incentive features of the labor market and thus the group coordination failure that characterizes many statistical discrimination equilibria.}. After all, the mistaken inference of unequal group ability levels indeed appears to be the origin of firms' inequitable hiring decisions. Moreover, when investment level is positively correlated with likelihood of being qualified, hiring based solely on investments is both rational and individually-fair. However, this group-blind solution fails to take into account a critical aspect of workers' investments---namely that they are \emph{choices} rather than \emph{givens}. Failure to recognize the upstream causes of observed data features brings to light the prickly notion of ``ground truth'' that has, from the start, plagued work on machine learning bias. Within a system as complex as the labor market, an input-output account of fairness that assesses the mapping of workers' investment levels to their hiring outcomes does not resolve the underlying source of inequalities that drives the differences in attributes between groups. Because both statistical discrimination and machine learning rely on data that harbor historical inequalities, \textit{local} fairness checks are often incapable of addressing the self-perpetuating nature of biases. Even without group biases, the paradox remains: the cyclic equilibrium ensures local procedural fairness---fairness with respect to investment choices---while maintaining global disparate outcomes.

The difficulty in pinpointing a particular cause of observed system-wide asymmetric outcomes challenges our mission in designing constraints to ensure fairness within the domain. If the outcomes themselves are trapped in a feedback loop, a successful fairness constraint should first jolt the system out of its current steady-state, and second, launch it on a path towards a preferable equilibrium. As such, a successful approach must consider fairness \textit{in situ.} This paper presents a \textit{domain-specific} dynamic model with an intervention that effects \textit{system-wide} impact, guaranteeing a group-equitable equilibrium that is stable and self-sustaining. 


In our model, workers invest in human capital, enter first a Temporary Labor Market (TLM) and then transition into a Permanent Labor Market (PLM)\footnote{Contracting in a segmented market is common in the labor economics literature. Of these, our work is most similar to Kim \& Loury \cite{loury2014collective}, but notably they model the effects of statistical discrimination, while ours explicitly requires group-equitable outcomes.}. We use this partition to impose a constraint on TLM hiring practices that enforces group statistical parity representation. However, the restriction need not apply in the PLM where firms select natural best response hiring strategies. Our employment model is \textit{reputational}---a worker carries an individual reputation, which is a summary of her past job performances and belongs to a group with a collective reputation, which is a measure of the proportion of its members producing ``good'' outcomes. 

Working within this model, we show that by imposing this constraint on firms' hiring strategies in the TLM, the resulting steady-state in the PLM is symmetric such that an equal proportion of workers in the two groups produce good outcomes and are thus hired. The labor market at equilibrium, both procedurally and in outcomes, satisfies leading notions of ``fairness''--group, individual, meritocratic \cite{feldman2015certifying, dwork2012fairness, kearns2017meritocratic}---discussed in the algorithmic fairness literature. Furthermore, we show that under particular labor market conditions, it Pareto-dominates the asymmetric outcomes that arise under two unconstrained rational hiring strategies: group-blind hiring and statistical discriminatory hiring. Our fairness intervention exploits the complementary nature of individual and collective reputations such that the system produces its own feedback loop that incrementally addresses initial inequalities in group social standing. As such, the TLM intervention need not be permanent---statistical parity of hired workers becomes the natural result of firms' optimal hiring strategies once group equality is restored and the fairness constraint becomes obsolete.




This paper's constraint-based approach to achieving equitable group outcomes in a reputational model of labor market interactions melds the perspectives and techniques of labor economics with the motivations of algorithmic fairness. However, our system-wide view also challenges a thread of work in the literature that characterizes notions of fairness as input-output-based properties of a decision-making function. By casting workers and firms as strategic agents in a dynamic game, we incorporate complexities of the labor market dynamic such as agents' expectations, incentives, and externalities that are otherwise difficult to encapsulate in a static classification setting. We advocate for an intervention that addresses the root of disparities between black and white workers' positions in the labor market and society---not only positions of unequal prospects and outcomes but as important, positions of unequal opportunities and, as a result, qualifications. Ensuring procedural fairness in the hiring decision alone is insufficient for this greater task. Our proposed constraint is designed to perturb a labor market at asymmetric equilibrium by co-opting the system's own cyclic effects to install group-equality that is self-sustaining in the long-term. 


In Section 2, we present a standard model of labor market dynamics and introduce our fairness intervention. Section 3 contains an overview of the equilibria results of the constrained-hiring model along with a comparison against equilibria arising from two rational hiring strategies free from such a constraint. The paper ends with a reflection on the equilibrium tendencies of discrimination and their implications on the design of fairness constraints. We also offer some comments on the dynamic feedback effects that are inherent features of persistent inequalities and the challenges they issue upon future work in algorithmic fairness.

\subsection{Related Work}

Within the algorithmic fairness literature, Zemel et al. \cite{zemel2013learning} address group and individual notions of fairness by constructing a mapping of agent data to an intermediate layer of clusters that each preserve statistical parity while obfuscating protected attributes. A second map taking cluster assignments to their final classifications then allows ``similar'' agents to be treated similarly. This dual-map approach roughly corresponds to the roles of the TLM and PLM in our model. Related work has sought distance metrics to guide the initial mapping \cite{dwork2012fairness}, but since criteria for similarity vary by domain, general approaches often face obstacles of application. Our paper's concentrated treatment of labor market dynamics aims to addresses this concern. We answer a call by Friedler et al. \cite{friedler2016possibility} to specify a particular world view of fairness within a domain and classification task. Our model starts with an assumption of inherent equality between groups. As such, differences in observable investment decisions or job outcomes are due to unequal societal standing, producing secondary effects of inequality, rather than fundamental differences in the nature of the individuals.

Labor market discrimination has been of long-standing interest in economics due to the persistent inequalities in employment prospects among groups of different race, gender, and other socially-salient attributes \cite{cain1986economic, altonji1999race, fryer2013racial}. Since most explicit forms of wage discrimination are now illegal in the U.S. and genetic accounts of group differences have been largely discredited \cite{neal1996role}, modern theories of labor market discrimination have updated the classical works---Becker's ``taste-based'' discrimination \cite{becker1971economics} and Phelps' model of exogenous group productivity differences \cite{phelps1972statistical}---by examining the \emph{social} sources of asymmetric outcomes. Research in the field has produced models that consider temporal dynamics, utilize distinct group cost functions, and develop wages endogenously \cite{chaudhuri2008statistical, antonovics2006statistical}. We follow in this line of work by incorporating a dynamic group reputation parameter into an individual's cost function, a modeling choice informed by the vast empirical literature showing the differential externalities produced by groups of differential social standing. Our model is not the first that makes explicit this linkage. In research examining the impact of neighborhood segregation on agents' accesses to resources for skill acquisition, Bowles, Loury, and Sethi \cite{bowles2014group} include a group ``skill share'' metric that functions similarly to our notion of group reputation in its effect on individuals' costs.


This paper also frames the hiring process as reputational in nature, following a distinct literature on collective reputation \cite{tirole1996theory,winfree2005collective}. Of these, our work shares most in common with the model proposed by Levin \cite{levin2009dynamics}, in which workers carry an individual reputation that contributes to their group's reputation. Levin shows that even when cost conditions evolve stochastically, reputations can produce a persistent feedback effect that leads to convergence to an asymmetric equilibrium in which groups occupy distinct social standings. Unlike in Levin, the notion of collective reputation in our model bears not only on workers' forward-looking expectations and incentives but also explicitly impacts future generations' investment costs. Additionally, since our work has in mind the information-processing capabilities of artificial intelligence agents, we formalize the concept of ``individual reputation'' as composed of a total history of previous outcomes. These additional ``data,'' while potentially overwhelming for human decision-makers, can be handled by an algorithmic decision-maker. Since the functionality of machine learning in the hiring process is ultimately based in a form of ``rational'' statistical discrimination of worker data and job histories, this strand of economics literature is particularly relevant for considerations of algorithmic fairness in the labor market.

\section{Model}
We highlight the role of the fairness constraint within the rest of the standard labor market dynamics of the model by utilizing a dual labor market setup composed of a Temporary Labor Market (TLM) and a Permanent Labor Market (PLM). In the former, a hiring constraint is established to ensure statistical parity, and in the latter, firms hire according to their best response hiring practices in a reputational model applied to the particular setting of employment. 
 
This partition does little to impinge upon the standard dynamics  of the labor market---workers flow from the TLM to the PLM, wages are labor-market-wide, and individual worker reputations in the PLM produce externalities for the collective group reputations that play a key role in individuals' pre-TLM investment decisions. 
\subsection{General Setup}
Consider a society of $n$ workers who pass through the labor market sequentially at times $t = 0, 1, ...$. The labor markets maintain a constant relative size: $m$ proportion of the workers reside in the TLM, and $1-m$ reside in the PLM. Movement is governed by Poisson processes---workers immediately replace departing ones in the TLM, transition from the TLM to the PLM according to the parameter $\kappa$, and leave the PLM at rate $\lambda$. 

Each worker belongs to one of two groups $\mu \in \{B, W\}$ with population share $\sigma_B$ and $1-\sigma_B$ respectively. We assume that these subpopulation proportions of workers are stable such that a worker of group $\mu$ who leaves the labor market is replaced via the birth of a new worker of the same group. 
The distribution of individual abilities, described by the CDF $F(\theta)$, is stable over time and identical across groups. In contrast, societal reputation varies with time and by group. A group's time $t$ reputation $\pi^\mu_{t}$ gives the proportion of all individuals in group $\mu$ who are producing ``good'' outcomes in the labor market, over the interval timespan $[t-\tau,t]$, where the parameter $\tau \ge 0$ controls the time-lag effect of a group's previous generations' performance on its present reputation. 

Prior to entering the labor market, workers select education investment levels $\eta$, weighing the cost of investment with its expected reward. Firms hire and pay workers based on expected performance, awarding wage $w_(g_t)$ for a ``good'' worker, where $g_t$ gives the proportion of ``good'' workers in the PLM at time $t$. To prevent constant fluctuation at each time step, the wage $w_{t} = w(g_{t'})$ updates in a Poisson manner such that $t' < t$ gives the time of the last wage change. The hiring process is formalized by assigning workers to either skilled or unskilled tasks with distinct wages. For simplicity, workers who do not pass particular hiring thresholds may still be considered ``hired,'' but they are assigned to an unskilled task and paid a wage normalized to $0$. As a function, the wage premium $w_t$ is decreasing in $g_t$, since as the relative supply of ``good'' workers increases, imperfect worker substitutability lowers their marginal productivity, thus decreasing wage. We impose a minimum wage $\underaccent{\bar}{w}$ such that $\lim_{g_t \to \infty} w(g_t) = \underaccent{\bar}{w}$ and a maximum wage $\bar{w}$ such that $\lim_{g_t \to 0} w(g_t) = \bar{w}$. In the context of the model, minimum and maximum wages should not be considered as only products of labor laws, rather they also act to track the supply of ``good'' workers relative to firms' demand. 


\subsection{Temporary Labor Market}
A worker $i$ of group $\mu$ chooses to invest in human capital $\eta_i \ge 0$ according to her expected wage gain of being in the skilled labor market $w_t$\footnote{Workers are boundedly rational and unable to anticipate future wage dynamics.} and her personal cost function for investment, $c_{\pi^{\mu}_{t}}(\theta_i, \eta_i)$, which is a function decreasing in her individual ability $\theta_i$ and increasing in her selected level of investment $\eta_i$. The incorporation of group reputation $\pi^\mu$ into an individual's cost function reflects the differential externalities produced by groups of differential social standing \citep{bowles2014group}. We posit that a worker belonging to a group with a superior societal reputation has improved cost conditions relative to her counterparts with equal ability in the lower reputation group. Formally, $\forall \pi^\mu_{t} < \pi^\nu_{t}, c_{ \pi^\mu_{t}}(\theta_i, \eta_i)$ is a positive monotonic transformation of $c_{\pi^{\nu}_{t}}(\theta_i, \eta_i)$. 

Investment in human capital operates as an imperfect signal, and workers have a hidden true type: \emph{qualified} or \emph{unqualified}, $\rho \in \{Q, U\}$. Let $\gamma: \mathbb{R}_{\ge 0} \rightarrow [0,1]$ be a monotonically increasing function that maps a worker's investment level to her probability of being qualified. Unlike in Spence's original work on education signaling \cite{spence1973job} in which investment confers no productivity benefits and thus operates purely as a signal to employers, in our model, a worker's chosen investment level $\eta$ has intrinsic value insofar as it is positively correlated with her likelihood of being qualified $\gamma(\eta)$. 

Given this setup, a firm's \textit{TLM hiring strategy} is a mapping $\mathcal{H}_T: \mathbb{R}_{\ge0}$ $\bigtimes$ $\mu \rightarrow \{0,1\}$ such that the hiring decision for worker $i$ is based only her observable investment level $\eta_i \in \mathbb{R}_{\ge 0}$ and group membership $\mu$. A worker who is hired into the TLM enters the pipeline and is eligible to compete for a PLM skilled job; a worker who does not pass the TLM hiring stage remains in the market but is permanently excluded from candidacy for the skilled wage. In this paper, we mainly consider only those workers who successfully enter the skilled hiring pipeline, considering all others as ``not hired.'' As such, we use the terms ``skilled'' and ``hired'' interchangeably. 





 
\begin{figure}
\centering
\includegraphics[scale=0.25, trim={1cm 13cm 1cm 7.2cm}, clip]{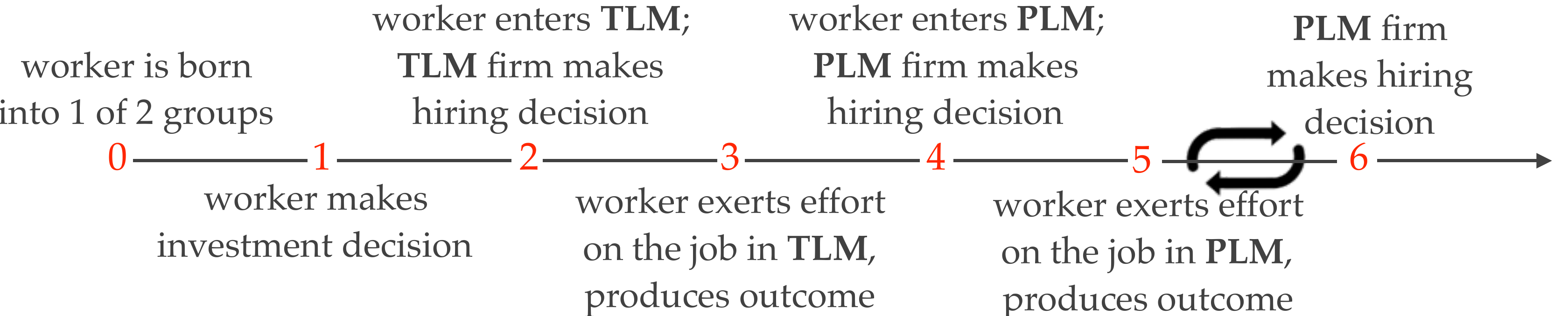}
\caption{Timeline of worker and firm interactions throughout the labor market pipeline.}
\label{fig:timeline}
\end{figure}

\subsection{Permanent Labor Market}
Labor market dynamics follow in the style of repeated principal-agent interactions with hidden actions (effort exertion) but observable histories (reputation of outcomes). Once hired into the TLM, a worker $i$ exerts on-the-job effort---choosing either \emph{high} ($H$) or \emph{low} ($L$) effort---which stochastically produces an observable \emph{good} ($G$) or \emph{bad} ($B$) outcome that affects her individual reputation and thus future reward. Exerting $L$ is free, but exerting $H$ bears cost $e_\rho(\theta_i)$, which is a function of qualification $\rho \in \{Q,U\}$ and ability level $\theta_i$. Effort is more costly for unqualified individuals: $\forall \theta_i, e_U(\theta_i) > e_Q(\theta_i)$. We emphasize here that the notions of ability level $\theta$ and qualification status $\rho$ are distinct worker qualities. A high ability worker is one who has the general attributes that bear on success in the realms of education and work, whereas a qualified worker is one who has the appropriate training and skills for a given job. We may say, very crudely, that a worker is ``born'' with an ability level and ``earns'' a qualification status. In our model, a worker's ability level precedes her investment decision, which begets a qualification status.

High effort increases the probability of a good outcome $G$. If $p_{\rho, k}$ gives the probability of achieving outcome $G$ with qualifications $\rho$ and effort level $k$, then the following inequalities hold.
\begin{align*}
p_{Q,H} > p_{Q,L}; p_{U,H} > p_{U,L}; p_{Q,L} > p_{U,L}
\end{align*}
Since the effect of qualifications on exerting high effort is already incorporated in its cost, $p_{Q,H}=p_{U,H}$, we write both quantities as $p_H$. We then simplify $p_{Q,L}$ and $p_{U,L}$ to $p_Q$ and $p_U$ respectively.  

We emphasize the distinction between the \textit{effort} exertion cost functions $e(\cdot)$ here and the previous \textit{investment} cost functions $c(\cdot)$---the former are pertinent to workers already in the labor market and differ by qualification status, whereas the latter relate to pre-labor-market decisions and differ by group membership. Separate cost functions allow for a finer analysis of the salient factors that influence agent behavior at distinct points of the labor market pipeline. The inclusion of group membership into human-capital investment costs reflects the genuine differences in resources available to workers of different groups in their paths to education attainment\footnote{We do not claim that group membership ceases to be a relevant factor impacting agent behavior once workers are in the labor market, but we note that a worker's qualifications, or the extent to which her skill investment proved to be successful, becomes an overriding determinant. Insofar as education investment bears on qualification status, a worker's group membership continues to impact her labor market outcomes.}. 


A worker keeps the same TLM job until the Poisson process with parameter $\kappa$ selects her to move into the PLM, where at each time step, she cycles through jobs, exerting a chosen effort level, producing an observable outcome, and accumulating a history of past performances that includes her TLM outcome. At each time step, firms in the PLM want to hire all and only those workers who consistently exert effort. To do so, firms distill a worker's history of observable outcomes into her ``individual reputation'' $\Pi^t_i$, which gives the proportion of outcomes $G$ in her recent length-$t$ history. In a labor market system of repeated worker-firm contracting, firms have the power to use these observable individual reputations to set self-enforcing relational contracts. A firm's \textit{PLM hiring strategy} is a mapping $\mathcal{H}_P: [0,1] \rightarrow \{0,1\}$ such that the decision is solely a function of $\Pi^t_i$. Figure \ref{fig:timeline} depicts a timeline of how workers move through the labor market pipeline and interact with firms. 

While ``fairness'' is a notoriously thorny ethical concept to define, the goal here of achieving long-term fairness is equivalent to attaining group equality in labor market outcomes. Since groups do not differ in fundamental or intrinsic ways, their job and wage prospects should also not systematically diverge at a fair steady-state. 



\begin{table}[ht]
\small
\centering
\caption{Table of notation}
\label{notation}
\begin{tabular}{l|l}
Notation & Significance \\ \hline
$F(\theta)$ & CDF of ability levels $\theta$ \\
$\pi^\mu$ & group $\mu$ reputation \\
$\sigma_\mu$ & group $\mu$ population share \\
$w_t$ & wage at time $t$ \\
$g^\mu_t$ & \begin{tabular}[c]{@{}l@{}}proportion of group $\mu$ workers\\ producing good outcomes at time $t$\end{tabular} \\
$\eta$ & investment level \\
$p_H, p_Q, p_U$ & probability of producing $G$ given effort level\\
$c_{\pi^\mu_t}(\theta, \eta)$ & cost of investment \\
$\gamma(\eta)$ & probability of being qualified \\
$\rho \in \{Q, U\}$ & hidden qualification status \\
$e_\rho(\theta)$ & cost of effort exertion \\
$\Pi^t_i$ & individual reputation at time $t$
\end{tabular}
\end{table}
\section{Results}
Reputation-based labor market models, such as the one described in this paper, can generate asymmetric group outcomes when firms utilize rational strategies such as statistical discrimination or group-blind hiring \cite{arrow1973theory,coate1993will,antonovics2006statistical,chaudhuri2008statistical}. Since this paper examines the effect of our proposed intervention on system-wide dynamics and outcomes, in the following section, we consider only those strategies and equilibria outcomes that arise in this fairness-constrained setting.   

\subsection{Equilibrium Strategies and Steady-States}
We start by describing TLM strategies resulting from the fairness constraint, then move onto the PLM and analyze firms' and workers' best response strategies together. Gameplay in the PLM mirrors repeated principal-agent interactions wherein firms have the power to enforce contracts by monitoring individual reputations, and thus we consider strategies that constitute a sequential equilibrium.

Since a firm in the TLM prefers candidates who are more likely to be qualified, optimal hiring follows a threshold strategy: Given a hiring threshold $\hat{\eta}$, $\forall i$ such that $\eta_i \ge \hat{\eta}$, $\mathcal{H_T}(i) = 1$, and inversely, $\forall i$ such that $\eta_i < \hat{\eta}$, $\mathcal{H_T}(i) = 0$. However, since firms must abide by the statistical parity hiring rule, their optimal threshold strategy is uniquely determined: if a firm aims to hire a fraction $\ell$ of all workers, its investment thresholds will be implicitly defined and group-specific, so that in the TLM, skilled employees from groups $\mu$ and $\nu$ will constitute $\sigma_\mu \ell$ and $(1-\sigma_\mu)\ell$ proportions of the full worker population respectively. 

A worker of group $\mu$, observing her group-specific TLM investment threshold $\widehat{\eta_\mu}$, will weigh her cost of investment with her expected wage gain $w_t$. All workers $i$ with $c_{\pi^{\mu}_{t}}(\theta_i, \widehat{\eta_\mu}) \le w_t$ will choose to invest exactly at the level $\eta_i = \widehat{\eta_\mu}$ and be hired for the skilled position in the TLM; all other workers will invest at level $\eta_i = 0$ and fail to enter the pipeline to compete for the skilled job. Workers who pass the first hiring stage know that their future PLM opportunities will depend on their observable outcome in the TLM, and as such they exert effort in a one-shot game. A worker $i$ with qualification status $\rho$ exerts high effort on the job if and only if $e_\rho(\theta_i) \le w_t (p_H-p_\rho)$.

As previously shown, while the statistical parity constraint preserves the fundamental equality of ability distributions $F(\theta)$ between groups, the group-specific investment thresholds $\widehat{\eta}_\mu$ generate group-specific investment strategies. As consequence, since investment has positive returns on qualification status, groups may have differing proportions of qualified candidates in the PLM pool. We denote by $\gamma^\mu_t$ the proportion of candidates in group $\mu$ who are qualified at time $t$, leaving $1-\gamma^\mu_t$ who are unqualified. Then the proportion of group $\mu$ workers in the TLM who produce good outcomes follows the recursive model
\begin{align}
{g^\mu_{t}} = & p_H[1-F(\widehat{\theta_Q})\gamma^\mu_t-F(\widehat{\theta_U})(1-\gamma^\mu_t)]+ p_{Q}F(\widehat{\theta_Q})\gamma^\mu_t \label{sseq}
\\
&+p_{U}F(\widehat{\theta_U})(1-\gamma^\mu_t) \nonumber
\\
&\text{where } \widehat{\theta_\rho} = e_\rho^{-1}(w_t(p_H - p_{\rho})) \nonumber
\\
& \text{ and } g_{t'} = \sigma_\mu \ell g^\mu_{t'} + (1-\sigma_\mu)\ell g^\nu_{t'} \nonumber
\end{align}
with $w_t = w(g_{t'})$ where $t'$ gives the time of the last wage update.



It is important to note that $g^\mu_t$ gives the proportion of workers in the skilled labor market who at time $t$ are producing good outcomes in their jobs. This quantity does not exactly coincide with group reputation, $\pi^\mu_t$, which gives a (time-interval average) normalized metric that scales with the proportion of \textit{all} members in group $\mu$--including those who are not granted entry into the skilled job pipeline--who are producing good outcomes. 

A PLM worker's future-anticipatory strategy is a selection of time, reputation, wage, and hiring threshold-dependent probabilities of effort exertion $\epsilon(\Pi_i^{t'})$ with $\Pi_i^{t'} \in \{\Pi^{t'}\}$ where the index $i$ of $\Pi_i^{t'}$ denotes a particular individual reputation level in the set of all possible reputation levels $\{\Pi\}$ and $t'$ tracks the length of time that has passed since the last wage update.
Supposing that workers engage in $N$-depth reasoning where $N \gg t'$, this quantity may be computed via backward induction on the continuation value for a given individual reputation, $V(\Pi^{t'}_i)$. With this setup, the continuation value $V(\Pi^N) = 0$, and the agent with ability $\theta $ and qualification $\rho$ solves the following dynamic programming problem
\begin{align}
&V(\Pi^{t'}_i, \hat{\Pi}^{t'}, w_t) = \sup_{\epsilon(\Pi^{t'}_i) \in [0,1]} \Big\{(1- \lambda)[V(\Pi^{t'+1}_i, G) [\epsilon(\Pi^{t'}_i)(p_H-p_\rho) + p_\rho]+ \nonumber 
\\
&V(\Pi^{t'+1}_i,B)[(1-\epsilon(\Pi^{t'}_i))(p_H-p_\rho)+1-p_\rho]] + \mathbbm{1}_{\Pi^{t'}_i \ge \hat{\Pi}^{t'}}w_t\Big\} \nonumber
\\
&\text{where } V(\Pi^{t'}_i, G) = V(\frac{\Pi^{t'}_i t'+1}{t'+1}, \hat{\Pi}^{t'}, w_t) \text{ and } V(\Pi^{t'}_i, B) = V(\frac{\Pi^{t'}_i t'}{t'+1},\hat{\Pi}^{t'}, w_t) \nonumber
\\
&\text{and } \forall t, w_t = w_T \text{ when the agent looks forward from time $T$} \nonumber
\end{align}
where the worker solves for optimal effort exertion probabilities $\epsilon(\Pi^{t'}_i)$ for each possible reputation $\Pi^{t'}_i \in \{\Pi^{t'}\}$, and high effort is only optimal at time $t$ if $V(\Pi^{t'}_i, G)(p_H-p_\rho) \ge e_\rho(\theta)$. 

If firms seek those workers who appear willing and able to exert high effort upon being hired, their equilibrium strategy is to select a reputation threshold $\hat{\Pi}^{t'} = p_H - \Delta_{t'}$ when facing a worker with history length $t'$ since the last wage update. $\Delta_{t'}>0$ acts as the firm's optimistic forgiveness buffer, permitting a worker's recent time $t'$ reputation to be slightly under the $p_H$ threshold, to ensure that it does not penalize workers who exert high effort but are unlucky and receive $B$ outcomes. An optimal choice of $\Delta_{t'}$ monotonically decreases in $t'$ toward $0$ as the reputation of a worker consistently exerting high effort converges to $p_H$ as $t' \rightarrow \infty$. Note that the firm must also take care not to decrease $\Delta_{t'}$ too slowly, lest workers are able to exert low effort and continue to be hired. Thus the firm optimizes its hiring threshold $\hat{\Pi}^{t'} = p_H - \Delta_{t'}$ by decreasing $\Delta$ just enough at each time step to motivate consistent high effort from workers who can afford it. All other workers exert low effort in each round. Thus given a firm's reputation threshold $\hat{\Pi}^{t'}$, its equilibrium PLM hiring strategy $\mathcal{H_P}$ is a mapping such that if and only if the worker's accumulated reputation since the last wage update $\Pi^{t'}_i$ exceeds the threshold $\hat{\Pi}^{t'}$, $\mathcal{H_P}(\Pi^{t'}_i) = 1$, and the worker is hired. Otherwise $\mathcal{H_P}(\Pi^{t'}_i) = 0$, and the worker does not earn the wage premium. This strategy is summarized in the following Proposition, and we defer the interested reader to the Appendix for its proof.

\begin{prop}
There exists a pair of PLM equilibrium strategies $(\mathcal{H}, \mathcal{E})$ of firm-hiring and worker-effort respectively such that 
\begin{enumerate}[(i)]
\item A firm's hiring strategy $\mathcal{H}$ is a selection of a reputation threshold function of the form $\hat{\Pi}^{t'} = p_H - \Delta_{t'}$, where $\Delta_{t'}$ is a monotonically decreasing function in ${t'}$, such that $\mathcal{H}(\Pi^{t'}) = 1$ if and only if $\Pi^{t'}_i \ge \hat{\Pi}^{t'}$, otherwise $\mathcal{H}(i) = 0$.
\item A worker's effort strategy $\mathcal{E}$ is a selection of effort levels that considers only the wage $w_t$ and cost of effort such that $\mathcal{E}(w_t) = H$ if and only if $e_\rho(\theta) \le w_t(p_H-p_\rho)$, else $\mathcal{E}(w_t) = L$. 
\end{enumerate}
\end{prop}

Interestingly, the strategies employed in the repeated worker-firm interactions in the PLM generate a recursive relationship of the proportion of ``good'' workers for each group that mirrors the structure of (\ref{sseq}). PLM firms' stringent threshold reputation hiring strategy imposes the same type of ``pressure'' on workers at each round of employment as does the single-shot game in the TLM. In both labor markets, every outcome ``counts.''

Having elaborated upon the dynamics of both the TLM and PLM, we incorporate worker movement and combine the results to obtain a recursive relationship that governs the sequence of workers' performance results from an initial wage $w_0$. Note that the multiplicity of possible firm hiring strategies produces a multiplicity of dynamic paths of outcomes $\{(g^\mu_t, g^\nu_t)\}^\infty_0$ to steady-state, but given that in our model, firms are willing to hire only and all workers who consistently exert high effort, firm and worker equilibrium strategies are as described in Proposition 1, there is a unique sequence of group outcome pairs $(g^\mu_t, g^\nu_t)$ such that there exists a time $t = T$ with the property that $\forall t \ge T$, $(g^\mu_{T}, g^\nu_{T}) = (g^\mu_{t}, g^\nu_{t})$. 

\begin{theorem} 
Under the described labor market conditions in which $\ell$ proportion of workers gain entry into the TLM and firms abide by the statistical parity hiring constraint, the proportion of all workers in group $\mu$ producing good outcomes at time $t$, $g^\mu_t$ in the full labor market follows the recursive system
\begin{align}
g^\mu_{t+1} =& p_H[1-F({\theta_Q})\gamma^\mu_t-F({\theta_U})(1-\gamma^\mu_t)]+ p_{Q}F({\theta_Q})\gamma^\mu_t
\\
&+p_{U}F(\theta_U)(1-\gamma^\mu_t) \nonumber
\\
&\text{where } \pi^\mu_t = \frac{\sigma_\mu \ell}{\tau} \sum^{t}_{j=t-\tau} g^\mu_j,
\label{eq:social}
\\
&\gamma^\mu_t = \phi(\widehat{\eta_\mu}(\pi^\mu_t)),
\label{eq:feedback}
\\
&{\theta_\rho} = e_\rho^{-1}(w_t(p_H - p_{\rho})),
\\
&g_{t} = \sigma_\mu \ell g^\mu_{t} + (1-\sigma_\mu)\ell g^\nu_{t},
\end{align}
where $\phi$ and $\widehat{\eta_\mu}$ in Eq. \ref{eq:feedback} are monotonically increasing functions whose composition combines the labor market's reputational feedback effect with firms' TLM constrained group-investment thresholds. Then there exists a unique stable symmetric steady-state equilibrium and convergence time $T$, wherein $\tilde{\pi}^\mu_{t} = \tilde{\pi}^\nu_{t} = \tilde{\pi}, \forall t > T$, satisfying system-wide fairness, with a corresponding unique stable wage $\tilde{w}$.
\label{theorem1}
\end{theorem}
To understand why the existence of this unique stable symmetric equilibrium is guaranteed when TLM firms are bound to the statistical parity requirement, consider the two variables that affect a group $\mu$ worker $i$'s likelihood of producing a good outcome: her ability level $\theta_i$ and her probability of being qualified $P(Q|\hat{\eta}_\mu) = \gamma^\mu$. Since there are positive returns to investment, $\gamma^\mu$ is increasing in $\pi^\mu$: As her group $\mu$ social standing rises, cost conditions improve, and as a result, workers in future generations are more likely to be qualified. 
With the imposition of the TLM hiring constraint, firms recognize the groups' different costs of investment and hire in a manner that retains equality between the two groups' underlying ability distributions $F(\theta)$ within the labor market, which assures that the proportions of workers producing good outcomes in each group $g^\mu$ do not diverge within the skilled labor market pipeline. Moreover, the statistical parity hiring constraint requires that firms hire in a manner such that workers from a disadvantaged group $\mu$ are not inequitably blocked from entering the skilled labor market and always constitute $\sigma_\mu \ell$ of the TLM. As a result of maintaining both identical ability distributions $F(\theta)$ and proportional representation $\sigma_\mu$ in the TLM, statistical parity hiring ensures that as group outcomes in the skilled labor market converge, so do group reputations.
Thus, the $\gamma_t$-generated positive feedback loop that pushes towards diverging group outcomes is always constrained, allowing the natural reputational feedback on group investment cost functions $c_{\pi^\mu}$ to drive the convergence of group outcomes and thus group reputations to a single steady-state value. Importantly, throughout the path of $\{(g^\mu_t, g^\nu_t)\}$ outcomes toward this symmetric steady-state, the ``severity'' of the TLM fairness constraint on firms' hiring strategies continually slackens until it recedes into disuse. For a full exposition of the proof, see the Appendix. 

Under statistical parity hiring in the TLM, groups with unequal initial social standing will gradually approach the same reputation level according to time-lag $\tau$. 
The constraint has the effect of co-opting the ``self-confirming'' loop for group reputation improvement---collective reputation produces a positive externality, lowering individual group members' cost functions, thus improving investment conditions for future workers, further raising individual and group reputation. We point out that the empirically-validated link between group reputations and members' investment costs makes a TLM statistical parity constraint a more efficient means of addressing group inequalities than a similar intervention in the PLM. Since the TLM represents the entry point into the market, enforcing statistical parity at the onset ensures that lower reputation workers are not disproportionately excluded from the pipeline as a whole.

We next compare this steady-state under the TLM constraint with long-term outcomes of other rational hiring strategies that are not bound by any fairness constraints and show that under particular market conditions, the fair steady-state is Pareto-dominant. 

\subsection{Comparative Statics with Unconstrained Hiring Strategies}
In the absence of any constraint, firms are free to select any strategy that will maximize their probability of employing high-ability, qualified workers. Two such common strategies are group-blind, sometimes called ``meritocratic,'' and statistical discriminatory hiring. We provide an overview of each practice and then continue on to comparing their long-term equilibria outcomes with the symmetric steady-state that arises under our TLM hiring constraint.

Consider a \emph{group-blind} TLM hiring strategy that is individual-based, operating under an equal-treatment philosophy. Without considering agent group membership--- suppose again $\mu \in \{B, W \}$---the firm hires a proportion $\ell$ of workers by selecting a single investment level threshold $\tilde{\eta}$ for all workers, implicitly defined as
\begin{align*}
\ell = (1-\sigma_B)\Big(1-F(c_{\pi^W}^{-1}(\tilde{\eta}(p_H - p_\rho))\Big)+ \sigma_B\Big(1-F(c_{\pi^B}^{-1}(\tilde{\eta}(p_H-p_\rho))\Big)
\end{align*}
where $\sigma_B$ and $1-\sigma_B$ give the proportion of individuals in groups $B$ and $W$ respectively, and the function $c_{\pi^\mu}(\cdot)$ determines the group $\mu$ investment level. Pragmatically under this strategy, the firm will examine the broad distribution of all investment levels and select a threshold above which it is willing to employ workers. This strategy is also rationalized by the fact that the threshold $\tilde{\eta}$ maximizes the expected number of hired workers who are qualified.

An alternative class of firm hiring strategies employ \emph{statistical discrimination}, in which priors regarding a worker's observable attributes, such as group membership, are used to infer a particular individual's hidden attributes. In particular, if TLM firms hold priors $\xi_B$ and $\xi_W$ about the two groups' capabilities, upon observing an applicant's group $\mu$ and investment level $\eta$, they will update their beliefs of the prospective employee's qualifications according to: 
\begin{align*} 
P(Q | \mu, \eta) = \frac{p_Q(\eta) \xi_\mu}{p_Q(\eta) \xi_\mu + (1-\xi_\mu)p_U(\eta)}
\end{align*}
where $p_Q(\eta)$ and $p_U(\eta)$ give the probability of a qualified and unqualified worker having investment level $\eta$ respectively.

\begin{theorem}
In a PLM with unsaturated demand ($w = \bar{w}$) for skilled workers, the TLM constraint leads to a symmetric steady-state equilibrium that Pareto-dominates the asymmetric equilibria that arise under group-blind and statistical discriminatory hiring.
\label{theorem2}
\end{theorem}

We present an abbreviated exposition of the underlying factors that drive unconstrained hiring strategies to Pareto-dominated outcomes. For the full account of the proof, see the Appendix.

Group-blind hiring satisfies neither of the two key constrained hiring guarantees described in the proof explanation for Theorem \ref{theorem1}---namely, groups no longer share equal ability distributions $F(\theta)$ nor are they proportionally represented in the market according to their demographic shares $\sigma_\mu$. The violation of both of these criteria contribute to group reputation divergence and thus the existence of persistent asymmetric outcomes between groups. 

At the asymmetric steady-state, groups retain distinct investment costs that, under a group-blind investment threshold, generate group-specific ability level thresholds $\widetilde{\theta_B}$ and $\widetilde{\theta_W}$. If group reputation $\pi_B < \pi_W$, then these ability thresholds may be ranked with respect to the threshold $\bar{\theta}$ that arises under the fairness constraint: $\widetilde{\theta_W} < \bar{\theta} <  \widetilde{\theta_B}$. These hiring strategies inequitably bound the proportion of able and qualified workers in group $B$ who are eligible to compete for skilled jobs, leaving behind an untapped source of group $B$ individuals who would have otherwise been hired. Under PLM conditions in which demand for skilled workers is unsaturated and the wage $w(g_t) = \bar{w}$, workers in group $W$ who are barred from entering the labor market in the proposed fair regime are not hired at equilibrium under group-blind hiring anyway. With strictly better-off employment outcomes for group $B$ workers and no worse outcomes for group $W$ workers, the constrained-hiring equilibrium Pareto-dominates the group-blind hiring equilibrium. 

Similarly, statistical discriminatory hiring leads to group-specific ability thresholds and does not guarantee statistical parity. As Coate and Loury \cite{coate1993will} show, self-confirming asymmetric equilibria also exist under this regime, wherein lower investment levels within the group with lower social standing are justified by firms' more stringent hiring standards. These effects have consequences that mirror the Pareto-dominated results under group-blind hiring.

\section{Discussion}
Describing disparate outcomes in employment as caused by rational agent best response strategies suggests that the field of algorithmic fairness should consider the labor market's inherent dynamic setting in its approach to potential interventions. Fairness constraints that are conceived as isolated procedural checks have a limited capacity to install system-wide fairness that is self-sustaining and long-lasting. The problem of fairness in the labor market is fundamentally tied to historical factors. Within nearly all societal domains in which fairness is an issue, past and current social relations differentially impact subjects, producing distinct sets of resources, options, and opportunities that continue to mark agents' choices and outcomes today. Empirical evidence points to what economist and social theorist Glenn Loury has called ``development bias,'' in which black members of society have reduced chances of realizing their potential, as the greater source of racial inequality in welfare outcomes than discriminatory hiring \cite{loury2009anatomy}. This perspective challenges the notion that assuring ``individual fairness'' of the actual procedure of hiring should be the primary concern in assuring a labor market that is unbiased as a whole. 

Not only is the standard learning theory formulation of the problem, in which agent attributes are treated as \textit{a priori} givens, inadequate to attend to development bias, it also neglects the (arguably) meritocratic goals of the labor market. In economic settings, rewarding merit primarily serves an instrumental purpose---to incentivize investment and effort---rather than existing simply to pass along desert-based awards to candidates. Framing the problem as one of clustering or classification fails to understand the labor market as an incentive-oriented system. Fairness criteria that solely assess an algorithm's treatment of workers' qualifications similarly fall into the trap of viewing hiring decisions only as rewards to meritorious individuals without considering the incentive purposes of the reward system at-large. 

In contrast, a dynamic model recognizes the ripple effect of development bias in the past and calls for a fairness intervention with incentive features that carries momentum into the future. The labor market as a source of economic opportunity is an ideal setting for a notion of fairness that is oriented toward a future beyond the short timeline of firm hiring cycles. It is precisely our focus on steady-state outcomes that allows for this long-term conception of fairness. However, it should be noted that the employment outcomes along the path to the symmetric equilibrium are by no means guaranteed to satisfy any notions of fairness, neither individual nor group. But we claim that conceiving of fairness in this way---as a project that aims to achieve permanent societal group-egalitarianism---is an ambition that is not only a worthy goal in itself but also one that we show may be economically socially optimal. 

Our model of individual reputations as a sequence of previous outcomes in the PLM fits within the hiring regime today, in which employers have increased access to worker data. Since algorithms will be largely responsible for making sense of this historical data, future work should consider how systems that sift through a worker's history should be designed to determine when group membership-related considerations, such as the ones embedded in the TLM constraint proposed here, should be taken into account. As machine decision-makers are deployed increasingly throughout hiring processes, we must grapple with a long tradition of explicit and implicit human biases that have rendered the labor market prone to discriminatory practices. We hope that this work can suggest ways that algorithmic fairness interventions can shift these hiring strategies towards contributing to a better, fairer future. 

While this paper has shown that imposing the TLM hiring constraint ultimately leads to a group-symmetric outcome, we do not claim that ours is the only intervention able to produce such an equilibrium. The labor market pipeline in reality is an elaborate sequence of agent choices and social stages that is much more complex and heterogeneous than our model's pre-TLM, TLM, and PLM periods. The true space of possible policy interventions dwarfs those considered in this work. Interventions aimed at reducing the economic inequalities that exist between black and white communities have been implemented at a variety of junctures in the standard social pipeline, ranging from direct governmental subsidy programs for childhood education costs in high-poverty areas to private companies' attempts at diversifying hiring by partnering with historically black colleges. As such, there may exist a multiplicity of intervention-types that all ultimately lead to group-egalitarian outcomes. Further analysis of the costs and efficiencies associated with each of these regimes will produce a richer understanding of potential fairness interventions and their concomitant welfare effects. Insofar as work in labor market fairness ought to inspire action and policy in the real world, these open questions will require both theoretical and empirical attention.

\section*{Acknowledgements}
This work is supported in part by an NSF Graduate Research Fellowship and NSF Grant \#CCF-1718549. 

\section*{Appendix}
\textbf{Proof of Proposition 1}

We want to show that the firm-set reputation threshold $\hat{\Pi}^{t'} = p_H-\Delta_{t'}$, where $t'$ is the time since the last wage update, enforces a worker strategy of effort exertion akin to that of the one-shot game, in which a worker exerts high effort if she can afford to do so and low effort otherwise. The firm, by setting its reputation threshold $\approx p_H$, is correctly restricting its membership to workers who appear to be consistently exerting high effort. By the Law of Large Numbers, a worker's recent time $t'$ individual reputation $\Pi^{t'}_i \to p_H$ almost surely as $t' \to \infty$ as long as she continuously exerts high effort at each time step. Moreover, since the relationship between effort exertion and \textit{G} or \textit{B} outcomes can considered Bernoulli trials with $p=p_H$, we use the law of the iterated logarithm to bound individual good workers' reputational deviations away from the theoretical mean $p_H$ as $t$ increases and have that for all $t=\tau$,
\begin{align}
|\Pi^{\tau}-\hat{\Pi}^{\tau}| \le \sqrt{\tau^{-1}(2*0.25 \text{log log} \tau)} 
\end{align}
Rubinstein and Yaari \cite{rubinstein1983repeated} have shown that, for a similar setup of imperfect observability and moral hazard in repeated interactions between insurers and clients, the enforceability of the insurers' strategies is dependent on the choice of the forgiveness buffer sequence. In our case, as long as $\Delta_{\tau} > \sqrt{\tau^{-1}(0.5\text{log log} \tau)} $ and the sequence $\Delta_{t'} \to 0$ monotonically, the Rubinstein-Yaari result carries over into employment relationships, and workers will always exert high effort when they can afford to do so. Importantly, our scenario does differ from theirs in two ways: 1) Workers do not stay in the labor market for an infinite number of rounds, 2) A firm must pay the labor-market-wide wage upon hiring a worker and cannot unilaterally deviate from the set price. Since workers exit the market according to a Poisson parameter $\lambda$ and the wage premium $w_t = w(g_{t'}) > 0$ is set to always provide a higher payoff for a worker than failing to be hired at all (due to the normalization with respect to the unskilled job wage), the memoryless death process ensures that a worker $i$ with qualifications $\rho$ will always find it within her interest to pursue the skilled job as long as it is individually rational for her to do so, i.e. $e_\rho(\theta_i) \le w_t(p_H-p_\rho)$. 
\vspace{5pt}

\noindent \textbf{Proof of Theorem \ref{theorem1}}

The TLM hiring constraint effects two guarantees: 1) It retains the fundamental equality of groups' ability level distributions $F(\theta)$ within the labor market; 2) It results in statistical parity in the proportion of workers offered skilled jobs in the TLM. Since the instantaneous time $t$ contributions to groups' full population societal reputations $\pi^\mu$ are equivalent to $g^\mu$ up to the same constant factor ($\ell$ proportion who enter the TLM), showing that the $g^\mu$ values converge is sufficient to show that group reputations $\pi^\mu$ do as well. 

Consider $g_{t+1} = \xi(g_{t})$ as a self-mapping $\xi: X \rightarrow X$ where $X$ is the unit interval $[0,1]$. Groups $\mu$ and $\nu$ have the same functional form of $\xi$ differing only in a few particular parameters, which will be addressed in the decomposition of $\xi$ into two separate functions. Assuming the two groups begin with unequal societal reputations, we suppose that (without loss of generality) $\pi_\nu < \pi_\mu$. We want to show that regardless of initial values $\pi^\nu_0 < \pi^\mu_0$, hiring outcomes will converge to achieve equal group outcomes system-wide under labor market dynamics with the TLM fairness constraint.

Due to effect 1) of the TLM hiring constraint and the fact that both groups experience the same labor-market-wide wage $w(g_{t})$, the PLM ability thresholds $\widehat{\theta_Q}$ and $\widehat{\theta_U}$ are also equivalent across groups. Thus the difference between the $g^\mu_t$ and $g^\nu_t$ arises due to the different corresponding proportions of qualified workers $\gamma^\nu_t < \gamma^\mu_t$ at time $t$. As such, we construct the function $\phi$ as a mapping of $\gamma_t \in [0,1]$ to $g_{t+1} \in [0,1]$, such that $g_{t+1} = \phi(\gamma_t)$. The function $\phi$ is generic across the two groups, and group differences are entirely encoded in the distinct inputs $\gamma^\mu_t$ and $\gamma^\nu_t$. 

Let's call $g^\mu_{t+1} = \phi(\gamma^\mu_t)$ and $g^\nu_{t+1} = \phi(\gamma^\nu_t)$, where we treat $\gamma^\mu$ and $\gamma^\nu$ as distinct points of the mapping $\phi$. Then, we have
\begin{align}
g^\mu_{t+1} =& p_H[1-F(\widehat{\theta_Q})\gamma^\mu_t-F(\widehat{\theta_U})(1-\gamma_t^\mu)]+p_QF(\widehat{\theta_Q})\gamma^\mu_t\label{eq:recurse}\\&+p_UF(\widehat{\theta_U})(1-\gamma^\mu_t) \nonumber
\end{align}
The difference $|g^\mu_{t+1} - g^\nu_{t+1}|$ is thus equivalent to the following
\begin{align*}
|\phi(\gamma^\mu_t)- &\phi(\gamma^\nu_t)|= |-p_HF(\widehat{\theta_Q})(\gamma^\mu_t - \gamma^\nu_t)+p_HF(\widehat{\theta_U})(\gamma^\mu_t - \gamma^\nu_t) 
\\
&+p_QF(\widehat{\theta_Q})(\gamma^\mu_t - \gamma^\nu_t)-p_UF(\widehat{\theta_U})(\gamma^\mu_t - \gamma^\nu_t)|
\\
& = (\gamma^\mu_t - \gamma^\nu_t)|p_H[F(\widehat{\theta_U}) - F(\widehat{\theta_Q})]+p_QF(\widehat{\theta_Q})-p_UF(\widehat{\theta_U})|
\end{align*}
We rewrite the quantity inside the absolute value:
\begin{align*}
|\underbrace{F(\widehat{\theta_U})[p_H-p_U]}_{\in (0,1)}+\underbrace{F(\widehat{\theta_Q})[p_Q-p_H]}_{\in (-1, 0)}| = |\epsilon_t|<1
\end{align*}
Together, $|g^\mu_{t+1} - g^\nu_{t+1}| = |\phi(\gamma^\mu_t)- \phi(\gamma^\nu_t)| \le |\epsilon_t| (\gamma^\mu_t - \gamma^\nu_t)$, $\forall \gamma^\mu_{t}, \gamma^\nu_{t} \in [0,1]$, and with the bound on $\epsilon$, $\phi$ is a contraction mapping.

Since group reputation considers the proportion of \emph{all} members in a group who are producing good outcomes,  statistical parity also has the upshot that a particular instantaneous time $t$ group reputation $\pi^\mu$ exactly scales with $g^\mu$ as each group is proportionally represented within the labor market according to its population-wide demographic share, so we need only consider $g^\mu_t$ values to determine the feedback loop property of collective reputation $\pi^\mu_t$ and group cost functions $c_\mu$ and $c_\nu$. Thus, the mapping $\psi: X \rightarrow X$, which maps normalized $g_{t+1} \in X = [0,1]$ to $\gamma_{t+1} \in X = [0,1]$ such that $\gamma_{t+1} = \psi(g_{t+1})$, is a weakly contracting map. 

We can now rewrite the recursive system $g_{t+1} = \xi(g_t)$ as a composition: $g_{t+1} = \xi(g_t) = \phi(\psi(g_t))$, where we have shown that $\phi$ is a contraction and $\psi$ is a short map. Then their composition $\xi$, which represents the recursive self-map determining the evolution of group-wide employment outcomes, is also a contraction map.

Then by the Banach Fixed Point Theorem, there is a unique fixed-point $\tilde{g} = \xi(\tilde{g})$ such that all initial points $g_i \in [0,1]$ converge to $\tilde{g}$ via a sequence of applications of the recursive relation $\xi$ as in (\ref{eq:recurse}): For any two group reputations $\pi^\mu$ and $\pi^\nu$ corresponding to initial points $g^\mu_0$ and $g^\nu_0$, there exists a $T$ such that $\forall t > T, \pi^\mu_t = \pi^\nu_t = \tilde{\pi}$ (similarly with $g^\mu$). At equilibrium, there is a unique wage $\tilde{w}$ corresponding to $\tilde{g}$, and the system admits group fairness. \qed
\vspace{5pt}

\noindent \textbf{Proof of Theorem \ref{theorem2}}

To show that the contraction and convergence assured by statistical parity hiring is not guaranteed under group-blind hiring, note that when $\pi_B < \pi_W$, necessarily $1-F(\widetilde{\theta_B}) < 1-F(\widetilde{\theta_W})$, and the composition of workers granted entry into the TLM does not satisfy statistical parity. We call the proportion of workers in the TLM belonging to groups $B$ and $W$, $k^B$ and $1-k^B$ respectively. Similarly to the proof of Theorem \ref{theorem1}, we decompose $g^\mu_t$ into the feed-forward labor market flow effect and the feedback natural reputational effect. However, since $\gamma^\mu$ for the two groups are the same, and $F(\widetilde{\theta_\mu})$ values differ, we instead write labor market flow as a function of $F(\widetilde{\theta_\mu})$, call it $\phi$\footnote{Note that in this proof, we also assume that $\widetilde{\theta_B} < \widetilde{\theta_U}$, but the proof carries through in the exact same manner when this is not true.}.

Then $g^W_{t+1} = \phi(F(\widetilde{\theta_Q}))$ and $g^B_{t+1} = \phi(F(\widetilde{\theta_B}))$, and 
\begin{align*}
|\phi(F(\widetilde{\theta_Q)})-\phi(F(\widetilde{\theta_B}))| = (F(\widetilde{\theta_B}) - F(\widetilde{\theta_Q}))\gamma(p_H-p_Q)
\end{align*}
Since $\gamma(p_H-p_Q) < 1$, $\phi$ thus also contracts in the feed-forward mechanism, however the function only captures the proportional $g^\mu_t$ dynamics from the TLM into the PLM, which does not scale with group reputation $\pi^\mu$ since statistical parity is not guaranteed. Instead, under group-blind hiring, group reputation, which captures the proportion of \emph{all} workers in the group who are producing good outcomes in the skilled labor, is a function of $k^B$, or the bottleneck of group proportionality created by the group-blind investment threshold. Thus the particular time $t$ normalized group societal reputation $\pi^B_t \propto \frac{k^B g^B_t}{\sigma_B} < g^B_t$ and $\pi^W_t \propto \frac{(1-k^B)g^W_t}{1-\sigma_B} > g^W_t$, and as a result, $|\pi^W_t - \pi^B_t| > |g^W_{t+1} - g^B_{t+1}|$. Since the mapping from $g^\mu_t \to \pi^\mu_t$ is not a contraction, the reputation feedback is not guaranteed to contract either. The system may thus reach an asymmetric equilibrium in which groups $B$ and $W$ maintain distinct investment costs and equal group reputations are never recovered.
 
We now show that this asymmetric outcome is Pareto-dominated by the hiring constraint-produced symmetric steady-state when PLM firms' demand for workers is not saturated and $w(\widetilde{g_t}) = \bar{w}$. For the two groups, $B$ and $W$, group-blind hiring imposes a single investment threshold $\tilde{\eta}$ such that hired workers in both labor markets have the same probability of being qualified regardless of group membership: $\gamma^\mu = \gamma^\nu = \gamma$. Suppose group reputations are not equal as in the case of the group-blind asymmetric equilibrium just proven, then group-blind hiring results in effective ability thresholds that may be ranked with respect to the  threshold $\bar{\theta}$ under statistical parity hiring. If $\pi_B < \pi_W$, then $\widetilde{\theta_W} < \bar{\theta} <  \widetilde{\theta_B}$. Note that throughout the paper, it is assumed that not all workers in the TLM are able to be hired in the PLM; therefore the ability threshold for exerting on-the-job effort is greater than the ability threshold resulting from the investment threshold under statistical parity-constrained hiring: $\widehat{\theta_Q} > \bar{\theta}$. 

When $\widehat{\theta_Q} < \widetilde{\theta_B}$, then TLM group-blind hiring leaves behind high ability workers in group $B$ who would have otherwise been hired in the PLM. In particular, all qualified workers in group $B$ with ability level $\theta \in [\widehat{\theta_Q}, \widetilde{\theta_B})$ are only hired in the fairness constrained equilibrium; under group-blind hiring, they are barred from entering the TLM. This result accords with the vicious circle of the asymmetric equilibrium, since the reputation gap $|\pi^B_t - \pi^W_t|$ and consequently, differences in group investment costs are maintained.


Further, since $1-F_g( \widehat{\theta_\rho}) < 1-F_f( \widehat{\theta_\rho})$ where $F_g$ and $F_f$ are the ability CDFs under the group-blind and fair regime respectively, in a labor market that demands more workers yet cannot sustain a higher wage ($\tilde{w} = w$)\footnote{There are a variety of reasons why an association of firms that demand more workers would be unable or unwilling to raise its wage higher $\tilde{w}=w$: A higher wage may encourage lower ability workers to apply and exert effort, and in reality, probabilities of success $p_H$ may be variable according to ability; thus the firm may want to \textit{a priori} exclude such workers. Wage caps may also result from firm-firm collusion on price.}, firms strictly prefer the steady-state equilibrium under the fairness constraint. This is because the effective higher ability threshold for group $B$ under the group-blind TLM strategy is inefficient, leaving behind an untapped resource of skilled and qualified individuals in group $B$ who would have otherwise been hired in the PLM. Even those workers in group $W$ with ability level $\theta \in [\widetilde{\theta_W}, \widehat{\theta_Q})$ who are only allowed to enter the TLM in the group-blind regime do not fare better, since all such workers have ability level lower than the PLM reputation threshold and are not hired at equilibrium anyway. Thus since some workers in group $B$ are strictly better off and workers in group $W$ no worse off, the asymmetric equilibria under group-blind hiring is Pareto-dominated by the symmetric one of the fair case. 

The proof of this result for the statistical discriminatory hiring regime follows similarly. If $\xi_W > \xi_B$, then $P(Q | W, \eta) > P(Q | B, \eta)$, and the groups face different incentive compatibility constraints. Self-confirming asymmetric equilibria also exist under this regime \cite{coate1993will}, and using the same argument about lost efficiency due to inequitable ability thresholds in the TLM for group $B$, these equilibria are also Pareto-dominated by hiring that abides by statistical parity. 


\bibliographystyle{unsrtnat}
\balance
\bibliography{bibliography-fairness}

\end{document}